\documentclass{article}
\usepackage{appendix}
\usepackage{amsmath}
\usepackage[retainorgcmds]{IEEEtrantools}
\usepackage{graphicx}

\begin{document}
\title{Parametric Resonance in a dissipative system  \emph{\'a la} Kr\"{o}nig-Penney}
\author{Loris Ferrari \\ Department of Physics and Astronomy (DIFA) of the University \\via Irnerio, 46 - 40126, Bologna,Italy}
\maketitle
\begin{abstract}
The competition between parametric resonance (PR) and dissipation is studied in the damped Kr\"{o}nig-Penney model, with time-dependent dissipation rate $\gamma(t)$. In the classical case, it is shown that dissipation leaves just a finite number of PR-bands at most, suppressing those at higher frequencies. An analysis of the Lewis-Reisenfeld invariant $I(q,\:p,\:\rho)$ is performed, showing that, in the PR regime, the auxiliary function $\rho(t)$ can be chosen bounded or unbounded, depending on the initial conditions.\newline       
\textbf{Key words:} Non autonomous systems; Parametric Resonance. 
\end{abstract}

e-mail: loris.ferrari@unibo.it

\section{Introduction}
\label{intro}
In the wide field of non autonomous classical systems, the study of the harmonic oscillator with time-dependent \emph{frequency} adresses a specific line of research, denoted as \textquoteleft parametric resonance\textquoteright$\:$(PR). In contrast to the standard resonance, PR may lead to an unlimited increase of the energy system at long times, even in the presence of dissipation. The basic model is a harmonic oscillator\footnote{Here we limit the study to the \emph{linear} case, even though the non linear effects are relevant for the suppression of PR, as shown, for example in ref.~\cite{DRR}.}, whose frequency $\widetilde{\omega}(t)$ fluctuates in time periodically with period $\tau$. If the product $\tau\widetilde{\omega}(0)$ falls in suitable \textquoteleft bands\textquoteright, the oscillation amplitudes and the energy can start growing up exponentially in time. In principle, the exponential instability is produced also by an arbitrary \emph{disordered} fluctuation\footnote{This is due to what, in the context of one dimensional quantum systems, is denoted as \textquoteleft Anderson localization\textquoteright.}, a case that will be not considered in the present work. A contrasting effect is the energy dissipation, that may lower or totally suppress the rate of exponential increase of the energy. In the classical case, the question which effect is overhelming leads one to study the motion equation

\begin{equation}
\label{ECl}
q^{(2)}+\gamma(t) q^{(1)}+\omega^2_{0}[1-\delta(t)]q=0\:,
\end{equation}
\\
where $\gamma(t)>0$ is the time-dependent rate of dissipation, $|\delta(t)|<1$ and $f^{(n)}(t):=\mathrm{d}^nf/\mathrm{d}t^n$.

The aim of the present work is solving eq.n~\eqref{ECl} exactly, for piece-wise constant $\delta(t)$ and $\gamma(t)$ (Fig. 1), varying in phase with each other, a model that we call  \emph{\'a la} Kr\"{o}nig-Penney. Actually, the classical equation \eqref{ECl}, in the absence of dissipation ($\gamma(t)=0$), has the same form as a Kr\"{o}nig-Penney model (KPM), i.e. an eigenvalue Schr\"{o}dinger equation in one dimension, with piece-wise constant \textquoteleft potential energy\textquoteright$\:\omega_{0}^2\delta(t)/2$ ($\hbar=1$, particle mass $=1$) and \textquoteleft energy eigenvalue\textquoteright$\:\:\omega_{0}^2/2$ \cite{EIB}. As already shown in ref~\cite{Me}, the gaps of KPM (infinite in number), correspond to the PR bands of instability for eq.n~\eqref{ECl}. In Section \ref{CL} we show that the dissipation reduces the number of PR bands to a finite value at most, by suppressing those at higher initial frequencies $\widetilde{\omega}(0)$ of the oscillator.

\begin{figure}[htbp]
\begin{center}
\includegraphics[width=4in]{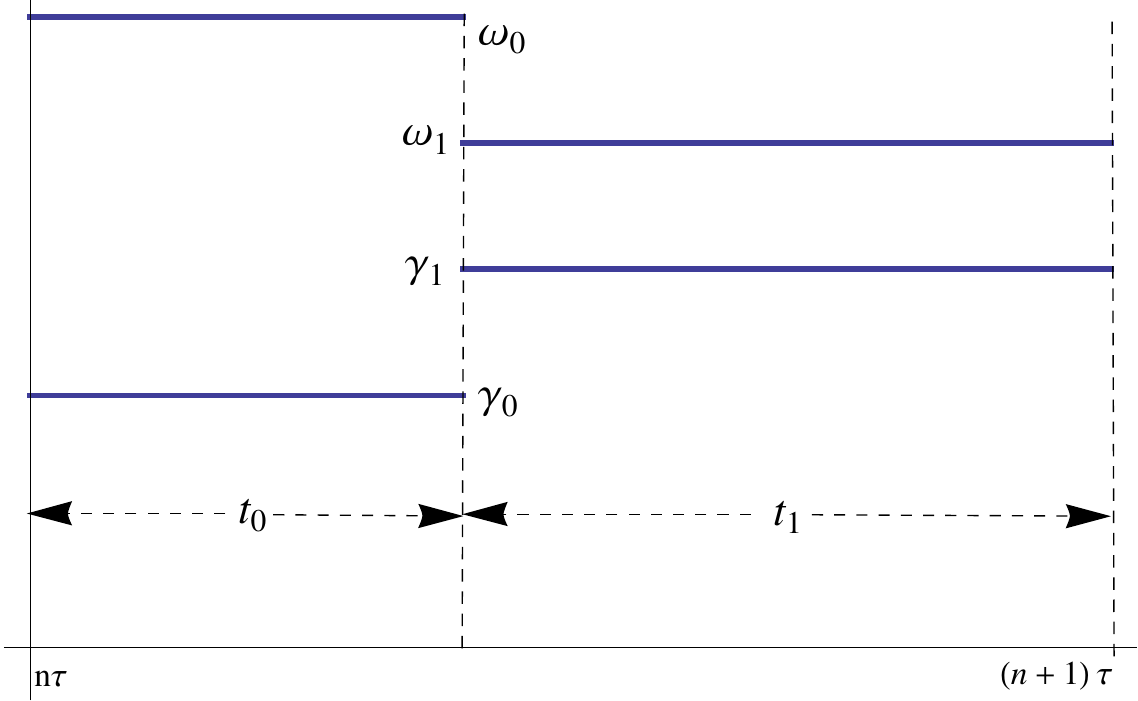}
\caption{\textbf{Kr\"{o}nig-Penney model for the time dependence of $\widetilde{\omega}$ and $\gamma$ in the $n$-t period}.}
\label{default}
\end{center}
\end{figure}

In Sections \ref{CL},  eq.n \eqref{ECl} is studied in detail, on noticing that eq.n~\eqref{ECl} follows by applying the Hamilton equations to the non autonomous Hamiltonian

\begin{subequations}
\label{H,omega}
\begin{equation}
\label{H(t)}
H(p,\:q,\:t)=\beta(t)\frac{p^2}{2}+\frac{\widetilde{\omega}^2(t)}{2\beta(t)}q^2,
\end{equation}
\\
with time-dependent frequency 

\begin{equation}
\label{omega(t)}
\widetilde{\omega}(t)=\omega_{0}\sqrt{1-\delta(t)}\:,
\end{equation}
\\
and damping

\begin{equation}
\label{beta(t)}
\beta(t)=\mathrm{exp}\left[-\int_0^t\gamma (t')\mathrm{d}t'\right]\:.
\end{equation}
\end{subequations}
\\ 
Since $H(t)$ is time-dependent, the research of other invariants of motion has a special interest. Section \ref{q^-3} deals with the Lewis-Reisenfeld invariant $I(q,\:p,\:\rho)$ of the classical motion \cite{LR}, expressed in terms of an auxiliary function $\rho(t)$. It is shown that in the PR regime the auxiliary function $\rho(t)$ can be chosen bounded ($I(q,\:p,\:\rho)=0) $ or unbounded ($I(q,\:p,\:\rho)>0$), depending on the initial conditions. 

\section{Parametric resonance \emph{vs} damping}
\label{CL}

For a KPM, as described by Fig. 1, the solution of eq.n~\eqref{ECl} can be written as:

\begin{subequations}
\label{q,Omega}
\begin{equation}
\label{q(t)}
q_\sigma(t)=A_\sigma(n)\mathrm{e}^{\Omega_\sigma t} +B_\sigma(n) \mathrm{e}^{\Omega_\sigma^{*}t}\:,
\end{equation}
\\    
with $\sigma=0$ for $t\in[n\tau,\:n\tau+t_0[$, and $\sigma=1$ for $t\in[n\tau+t_0,\:(n+1)\tau[$, $\tau =t_0+t_1$ being the period of the perturbation. The expressions for the $\Omega_\sigma$'s, resulting from eq.n~\eqref{ECl}, are: 

\begin{equation}
\label{Omegaj}
\Omega_\sigma=\mathrm{i}\overbrace{\omega_\sigma\left[\sqrt{1-\xi_\sigma^2}\right]}^{W_\sigma}-\frac{\gamma_\sigma}{2},\:\:\xi_\sigma:=\frac{\gamma_\sigma}{2\omega_\sigma},\:\:\omega_\sigma=\begin{cases}
\omega_0 & \sigma=0\\
\\
\omega_0\sqrt{1-\delta} & \sigma=1
\end{cases}
\end{equation}
\end{subequations}
\\
($\xi_\sigma^2<1$). We are interested in the overall change of the 2D vector

\begin{equation}
\vec{q}_1(\tau_n):=
\begin{vmatrix}
A_1(n)\mathrm{e}^{\Omega_0\tau_n}\\
\\
B_1(n) \mathrm{e}^{\Omega_0^{*}\tau_n}
\end{vmatrix}\:,
\end{equation}
\\
passing from $\tau_n=n\tau$ to $\tau_{n+1}$, which means studying the corresponding \emph{transfer matrix} $\mathcal{T}$, such that

\begin{equation}
\vec{q}_1(\tau_{n+1})=\mathcal{T}\:\vec{q}_1(\tau_{n})\:\Rightarrow\:
\begin{vmatrix}
A_1(n+1)\mathrm{e}^{\Omega_0\tau_{n+1}}\\
\\
B_1(n+1) \mathrm{e}^{\Omega_0^{*}\tau_{n+1}}
\end{vmatrix}
=\mathcal{T}
\begin{vmatrix}
A_1(n)\mathrm{e}^{\Omega_0\tau_{n}}\\
\\
B_1(n) \mathrm{e}^{\Omega_0^{*}\tau_{n}}\
\end{vmatrix}\:.
\end{equation}
\\    
Since the periodicity of the KPM yields:

\begin{equation}
\vec{q}_1(\tau_n)=(\mathcal{T})^{n}\vec{q}_1(0)\:,
\end{equation}
\\
the eigenvalue equation

\begin{equation}
\label{EigenT}
\mathcal{T}\:\vec{q}_\pm=\lambda_\pm\vec{q}_\pm
\end{equation}
\\
plays a crucial role. After projecting the initial vector $\vec{q}_1(0)$ on the eigenvectors $\vec{q}_\pm$ (eq.n~\eqref{EigenT}), the solution at $\tau_n$ takes the form 

\begin{equation}
q(\tau_n)=B_+\lambda_+^n + B_-\lambda_-^n\:, 
\end{equation}
\\
which determines the asymptotic behaviour, depending on the sign of the quantities $\ln |\lambda_\pm|$. In a PR band, the solution diverges exponentially, unless one picks the exact initial condition which projects the solution on the eigenstate $\vec{q}_\pm$, corresponding to $|\lambda_\pm|<1$. 

Once again, one can take advantge of the periodicity of KPM in the construction of the transfer matrix $\mathcal{T}$, which can be reduced to the initial period, from the continuity conditions:

\begin{align}
\vec{q}_0(t_0)&=\vec{q}_1(t_0)\:,\vec{q}_0^{\:(1)}(t_0)=\vec{q}_1^{\:(1)}(t_0)\nonumber\\
\label{contcond}\\
\vec{q}_0(t_0+t_1)&=\vec{q}_1(t_0+t_1)\:,\vec{q}_0^{\:(1)}(t_0+t_1)=\vec{q}_1^{\:(1)}(t_0+t_1)\:,\nonumber
\end{align}
\\   
leading to the following factorized form:

\begin{subequations}
\label{allTau}
\begin{equation}
\label{Tau}
\mathcal{T}=
\begin{vmatrix}
\mathcal{D} &  \mathcal{O}\\
\\
\mathcal{O}^* & \mathcal{D}^*   
\end{vmatrix}
= 
\begin{vmatrix}
\mathrm{e}^{\Omega_0(t_0+t_1)} & 0\\
\\
0 & \mathrm{e}^{\Omega_0^*(t_0+t_1)}
\end{vmatrix}
\mathcal{T}_1\mathcal{T}_0\:,
\end{equation}
\\
where the matrices

\begin{equation}
\label{Taualpha}
\mathcal{T}_\sigma=
\begin{vmatrix}
\mathcal{D}_\sigma &  \mathcal{O}_\sigma\\
\\
\mathcal{O}_\sigma^* & \mathcal{D}_\sigma^*   
\end{vmatrix} 
\end{equation}
\\
can be written in terms of the matrix elements:

\begin{align}
\label{D,O}
\mathcal{D}_\sigma&=\frac{\Omega_\sigma-\Omega_{1-\sigma}^*}{\Omega_{1-\sigma}-\Omega_{1-\sigma}^*}\exp\left[(\Omega_\sigma^*-\Omega_{1-\sigma})(t_\sigma+\sigma \:t_{1-\sigma})\right]\nonumber\\
&\quad\quad\quad\quad\quad\quad\quad\quad\quad\quad\quad\quad\quad\quad\quad\quad\quad\quad\quad\quad\quad(\sigma=0,\:1)\\
\quad\mathcal{O}_\sigma&=\frac{\Omega_\sigma^*-\Omega_{1-\sigma}}{\Omega_{1-\sigma}^*-\Omega_{1-\sigma}^*}\exp\left[(\Omega_\sigma-\Omega_{1-\sigma})(t_\sigma+\sigma \:t_{1-\sigma})\right]\nonumber
\end{align}
\end{subequations}
\\
and their complex conjugates. Since now on, we adopt the convention:
 \begin{equation}
z_0-z_1:=\Delta z\quad;\quad \langle\: z\:\rangle:=\frac{z_0+z_1}{2}\:,\label{Deltas}
\end{equation}
\\
for any pair of quantities $z_\sigma$, depending only on the index $\sigma=0,\:1$.
 
The eigenvalues $\lambda_\pm$ of $\mathcal{T}$ follow from the equations:

\begin{subequations}
\label{lambdaall}
\begin{equation}
\label{lambda1}
\lambda_+\lambda_-=\mathrm{det}\left[\mathcal{T}\right]=|\mathcal{D}|^2-|\mathcal{O}|^2
\end{equation}
\\
\begin{equation}
\label{lambda2}
\lambda_++\lambda_-=\mathrm{tr}\left[\mathcal{T}\right]=2\mathrm{Re}\left[\mathcal{D}\right]\:.
\end{equation}
\end{subequations}
\\
Recalling that $\tau=t_0+t_1$ represents the total period of the external perturbation leading to the time changes of $\widetilde{\omega}(t)$ and $\gamma(t)$, let one define:

\begin{subequations}
\label{gamma,thetas,A}
\label{definitions}
\begin{align}
\widetilde{\gamma}&=\omega_0\frac{\xi_0t_0+\xi_1t_1}{\tau}=\omega_0\tau\left(\langle\:\xi\:\rangle+\frac{\Delta\xi\Delta t}{2\tau}\right)\label{gamma}\\
\nonumber\\
\theta_\sigma&=W_\sigma t_\sigma\quad(\sigma=0,\:1,\text{recall \eqref{Omegaj}})\label{thetas}\\
\nonumber\\
A&=\frac{1-\xi_0^2+(1-\xi_1^2)(1-\delta)+\Delta\xi^2}{2\sqrt{(1-\xi_0^2)(1-\xi_1^2)(1-\delta)}}\label{A}.
\end{align}
\end{subequations}
\\
With some lenghthy but straightforward calculations, involving eq.ns~\eqref{allTau} and \eqref{Omegaj}, equations \eqref{lambdaall} become:

\begin{subequations}
\label{lambdaall2}
\begin{equation}
\lambda_+\lambda_-=\mathrm{e}^{-2\widetilde{\gamma} \tau}\:,
\end{equation}
\\
\begin{align}
\lambda_++\lambda_-=2\mathrm{e}^{-\widetilde{\gamma} \tau}\underbrace{\left[\cos\theta_0\cos\theta_1-A\:\sin\theta_0\sin\theta_1\right]}_{R}\:,
\end{align}
\\
which yields:

\begin{equation}
\label{lambda3}
\lambda_\pm=\left(R\pm\sqrt{R^2-1}\right)\mathrm{e}^{-\widetilde{\gamma} \tau}\:.
\end{equation}
\end{subequations}
\\
The necessary and sufficient condition for PR to overhelm dissipation is $|\lambda_+|$ or $|\lambda_-|$ larger than 1. Of course, a necessary condition is $R^2>1$, otherwise $|\lambda_\pm|=\mathrm{e}^{-\widetilde{\gamma} \tau}$. 

In view of concrete applications, it is convenient to define an appropriate \textquoteleft variable\textquoteright, and a set of fixed \textquoteleft parameters\textquoteright. Though the choice is arbitrary, a general aspect, common to all problems involving PR, is the quantity $\omega_0\tau$, expressing the ratio between the unperturbed oscillator frequency $\omega_0$ and the frequency of the external perturbation $2\pi/\tau$. Choosing $\omega_0\tau$ as the variable, the parameters to be kept fixed are: $\langle\:\xi\:\rangle$, $\Delta\xi$, $\delta$ and $\Delta t/\tau$ (recall eq.ns~\eqref{Deltas}, \eqref{Omegaj}). This, however, implies that $\gamma_\sigma\propto\omega_0$, i.e. that the dissipation is determined by the oscillations themselves, and that the time intervals $t_\sigma$ are proportional to $\tau$. Clearly, those assumptions are reasonable, but not mandatory. 

With this preliminary, the notion of \textquoteleft band\textquoteright$\:$ can be applied either to the unperturbed oscillator frequency $\omega_0$, with $\tau$ fixed, or, by inversion, to the frequency of the external perturbation $2\pi/\tau$, with $\omega_0$ fixed. It should be noticed that the limit of vanishing $\xi_\sigma$'s yields $\widetilde{\gamma}=0$ (eq.n~\eqref{gamma}). In this dissipationless case, the condition $R^2>1$ yields an \emph{infinite} sequence of PR-bands (Fig. 2a), corresponding to the energy \emph{gaps} of the standard quantum KPM. In contrast, the PR-bands at higher $\omega_0\tau$ are totally suppressed by dissipation, which leaves just a \emph{finite} number of bands where PR prevails (Fig. 2b). A further non trivial aspect that deserves attention is the asymmetry of the PR effects with respect to the sign of $\Delta\xi\Delta t$. In Fig.s 2b, 2c it is seen that the passage from negative to positive sign may suppress PR at all. Actually, the sign of $\Delta\xi\Delta t$ is determined by applying the larger ($+$) or the smaller ($-$) dissipation rate for the longer time interval between $t_0$ and $t_1$, which means simply to apply a larger or a smaller average dissipation rate, over the whole period $\tau$. 

\begin{figure}[htbp]
\begin{center}
\includegraphics[width=5in]{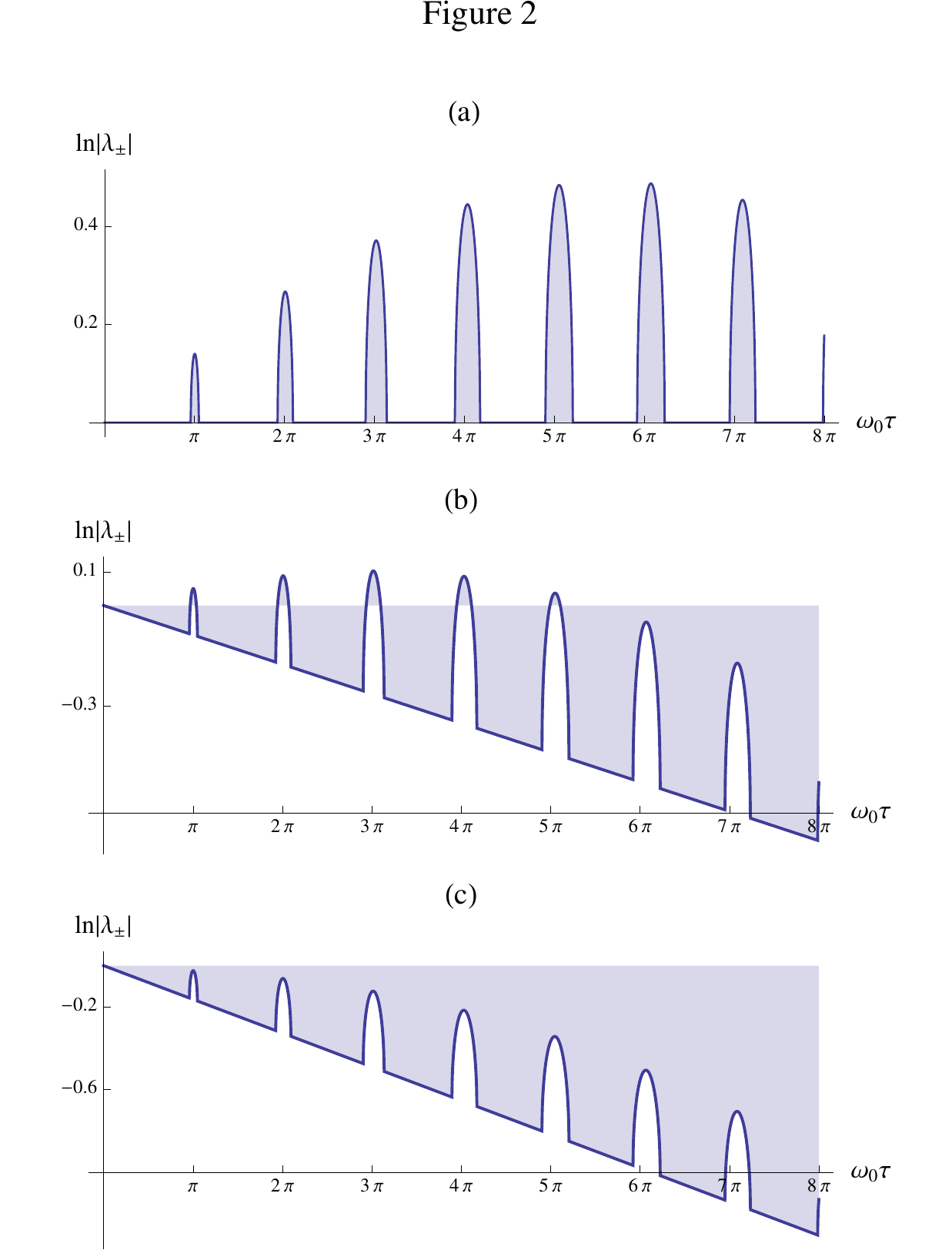}
\caption{\textbf{Parametric Resonance Bands}. The logarithm of $|\lambda_\pm|$ is plotted against $\omega_0\tau$ ($\delta=0.23$, $\langle\:\xi\:\rangle=0.04$, $|\Delta\xi|=0.03$, $|\Delta t/\tau|=0.8$). PR bands correspond to the filled regions where the logarithm is positive. \textbf{(a)}: infinite number of PR bands (no dissipation); \textbf{(b)}: suppression of high frequency PR bands by dissipation with $\Delta\xi\Delta t<0$; \textbf{(c)}: total suppression of PR bands by dissipation with $\Delta\xi\Delta t>0$. }
\label{default}
\end{center}
\end{figure}

It is useful recalling that $\delta$ refers to the time changes of the oscillator frequency, and \emph{may} lead to PR, under suitable conditions. In contrast, $\xi_{0,\:1}$ refer to the exponential damping, which \emph{certainly} leads to dissipation effects. Those parameters are dimensionless, so that the conditons $|\delta|<<1$ and $\xi_{0,\:1}<<1$ determine the regime of small fluctuations and dissipation. An analytic expression of what appears in Figure 2 can be obtained at the lowest order in $\delta$ and $\xi_{0,\:1}$, as shown in Appendix A. In particular, the expression 
   
\begin{equation}
\label{omegatau}
(\omega_0\tau)_n=\underbrace{n\pi\left(1+\frac{\delta\Delta t}{4\tau}\right)}_{PR-\text{band centre}}\quad(n=1,\:2,\:\cdots\:)
\end{equation}
\\
yields the centres of the PR-bands, which are the same as in the absence of dissipation, at the lowest order of approximation in $\delta$, $\xi_{0,1}$, and are infinite in number. The suppression of the frequency bands at higher $\omega_0\tau$, shown in Fig. 2, can be seen analytically by studying the PR-bandwidth, i.e. the size of the interval where $\omega_0\tau-(\omega_0\tau)_n$ can vary, in order that PR prevails. From the calculations developed in Appendix A, one gets, actually:

\begin{subequations}
\label{Bandwidth}
\begin{align}
&|\omega_0\tau-(\omega_0\tau)_n|<\nonumber\\
\nonumber\\
&<\underbrace{\left[\left(\frac{\delta^2}{4}+\Delta\xi^2\right)\frac{1-(-1)^n\cos^2(n\pi\Delta t/\tau)}{2}-(n\pi)^2\left(\langle\:\xi\:\rangle+\frac{\Delta\xi\Delta t}{2\tau}\right)^2\right]^{1/2}}_{\text{half } PR-\text{bandwidth}} \label{bandwidth}
\end{align}
\\
$(n=1,\:2,\:\cdots)$, provided that:

\begin{equation}
n\pi<\frac{\sqrt{\left(\delta^2+4\Delta\xi^2\right)\left[1-(-1)^n\cos^2(n\pi\Delta t/\tau)\right]}}{\sqrt{2}\left|2\langle\:\xi\:\rangle+\Delta\xi\Delta t/\tau\right|}\:.
\end{equation}
\end{subequations}
\\
Equations \eqref{Bandwidth} clearly show that the band index $n$ is upperly limited, for PR to overhelm dissipation. In addition, the effect of the asymmetry in the sign of $\Delta\xi\Delta t$ is clearly displayed too, since a negative or positive sign broadens or narrows down the bands, respectively.

It is useful recallig that, according to Appendix A, the eigenvalues $\lambda_+$ and $\lambda_-$ may be both responsible for PR, if the PR-band index $n$ is even or odd, respectively.

\section{The invariant $I$ and the auxiliary function $\rho(t)$.}
\label{q^-3}

Since a non autonomous Hamiltonian does not coincide, in general, with the system's energy, a preminent question is the energy evolution, in the presence of external sources and/or adsorbers. In particular, the classical energy corresponding to eq.n~\eqref{H(t)} reads:
 
\begin{subequations}
\begin{equation}
\label{Ecl}
E_{cl}\left(q,\:q^{(1)},\:t\right)=\frac{\left[q^{(1)}\right]^2+\widetilde{\omega}^2(t)q^2}{2}.
\end{equation}
\\
However, the Hamilton equations that follow from \eqref{H(t)} yield, in particular, $q^{(1)}=\beta p$, hence, recalling eq.n~\eqref{H,omega}, the Hamilton expression of the energy is\footnote{Here and in what follows it is useful to keep in mind the difference between the \textquoteleft true\textquoteright$\:$frequency $\widetilde{\omega}(t)$, with limited time-variations (Fig. 1), and $\omega(t)=\widetilde{\omega}(t)/\beta(t)$, with \emph{exponentially diverging} time variations.}: 

\begin{equation}
\label{Estrange}
\mathcal{E}(q,\:p,\:t)=\beta^2(t)\frac{\left[p^2+\omega^2(t)q^2\right]}{2}\:.
\end{equation}
\end{subequations}
\\

In the present case, neither the Hamiltonian $H(p,\:q,\:t)$, nor the energy $\mathcal{E}(q,\:p,\:t)$ are invariant. Therefore, the research of a motion constant has a special interest. It is well known that an invariant $I$ of the motion, resulting from eq.n~\eqref{ECl}, reads \cite{LR, PRA1992, PRA1997}\footnote{In ref. \cite{PRA1997} formula (4) for $I$ is to be corrected by replacing $(q/\rho)^{1/2}$ with $(q/\rho)^2$.}:

\begin{subequations}
\label{I,rho}
\begin{equation}
\label{I}
I=\frac{1}{2}\left[\left(\frac{q}{\rho}\right)^2+\left(p\rho-\frac{\rho^{(1)}q}{\beta}\right)^2\right]\:,
\end{equation}
\\
provided $q$ satisfies eq.n~\eqref{ECl} and the auxiliary function $\rho$ satisfies:

\begin{equation}
\label{eq.rho}
\rho^{(2)}+\gamma(t)\rho^{(1)}+\widetilde{\omega}^2(t)\rho-\frac{\beta^2(t)}{\rho^3}=0\:.
\end{equation}
\end{subequations}
\\
Hence, in the study of the classical non autonomous oscillator (and in the quantum case too \cite{LR, DEFL}), equation \eqref{eq.rho} plays a special role. First of all, it is immediately seen that $\rho(t)$ is constant in sign, since the repulsive force $\beta^2/\rho^3$ prevents $\rho$ to cross zero. Here and in what follows, $\rho$ will be assumed positive. The aim of the present section is to show that the auxiliary function $\rho(t)$, satisfying eq.n~\eqref{eq.rho} under PR conditions, can behave asymptotically in two different ways, depending on the invariant $I$ (eq.n~\eqref{I}). For $I=0$, $\rho$ can be chosen bounded and vanishing at long times, if dissipation does apply. If $I>0$, $\rho$ can be chosen to oscillate with diverging amplitudes. 

When applied to eq.n~\eqref{H(t)}, the Hamilton equations yield $p=q^{(1)}/\beta$, whence $q(t)=0\Rightarrow I=0$. In this case, one can find a \emph{bounded} solution of equation \eqref{eq.rho}, whose form is especially simple in the limit of small time changes of the parameters. On account of the definitions in eq.n~\eqref{Omegaj}), let:

\begin{equation}
\label{W(t)}
W(t):=\widetilde{\omega}(t)\sqrt{1-\xi^2(t)}=
\begin{cases}
W_0=\omega_0\sqrt{1-\xi^2_0}&\text{ for }n\tau\le t <n\tau+t_0\\
\\
W_1=\omega_1\sqrt{1-\xi^2_1}&\text{ for }n\tau+t_0\le t <(n+1)\tau
\end{cases}
\end{equation}
\\
($n=0,\:1,\:\cdots$). On setting

\begin{equation}
\label{rho2}
\rho(t)=\sqrt{\beta(t)}r(t)\:
\end{equation}
\\
in eq.n \eqref{eq.rho}, the motion equation for $r$ reads:

\begin{equation}
\label{eq.r}
r^{(2)}+\overbrace{W^2(t)r-\frac{1}{r^3}}^{\partial U(r,t)/\partial r}=0\:,
\end{equation}
\\
to be solved with suitable continuity conditions. Let $\widetilde{\omega}^*$ be an arbitrary value, between $\omega_0$ and $\omega_0\sqrt{1-\delta}$. Let, in addition: 

\begin{equation*}
r(t)=\frac{1}{\sqrt{\widetilde{\omega}^*}}+\Delta r(t)\:.
\end{equation*}
\\
Since  

\begin{equation*}
1/r^3(t)=(\widetilde{\omega}^*)^{3/2}-3(\widetilde{\omega}^*)^2\Delta r(t)+\mathrm{o}(\Delta r^2)\:,
\end{equation*}
\\
and $W(t)-\widetilde{\omega}^*=\mathrm{o}(\delta,\gamma)$ (eq.n~\eqref{W(t)}), equation~\eqref{eq.r} yields:

\begin{equation}
\label{Deltar}
\Delta r^{(2)}+\left[W^2(t)+3(\widetilde{\omega}^*)^2\right]\Delta r=\underbrace{(\widetilde{\omega}^*)^{3/2}-\frac{W^2(t)}{\sqrt{\widetilde{\omega}^*}}}_{\mathrm{o}(\delta(t),\:\xi(t))}+\mathrm{o}(\Delta r^2)\:,
\end{equation}
\\
which shows that $\Delta r(t)$, in the small oscillations limit in which $|\Delta r(t)|$ is smaller than or comparable to $1/\sqrt{\widetilde{\omega}^*}$, behaves like a harmonic oscillator with periodic driving $\mathrm{o}(\delta(t),\:\xi(t))$ and frequency $\sqrt{W^2(t)+3(\widetilde{\omega}^*)^2}$ (different from $\widetilde{\omega}(t)$). If $\sqrt{W^2(0)+3(\widetilde{\omega}^*)^2}$ does \emph{not} fall in a PR band, due to an appropriate choice of $\widetilde{\omega}^*$, the oscillations remain small for an arbitrary choice of the initial conditions on $\Delta r(t)$, such as, in particular:

\begin{subequations}
\begin{equation}
\label{in.cond.Delta}
\Delta r(0) =\frac{1}{\sqrt{\widetilde{\omega}_0}}-\frac{1}{\sqrt{\widetilde{\omega}^*}}\quad;\quad\Delta r^{(1)}(0)=\frac{\gamma_0}{2\sqrt{\omega_0}}\:, 
\end{equation}
\\
corresponding to 

\begin{equation}
\label{in.cond.rho}
\rho(0)=1/\sqrt{\omega_0}\quad;\quad\rho^{(1)}(0)=0
\end{equation}
\end{subequations}
\\
(eq.n~\eqref{rho2}). Hence, $r(t)$ can be chosen to oscillate about the minimum (in $r$) of the potential energy $U(r,t)$, corresponding to the total force in Fig.3. If so, the auxiliary function $\rho(t)$ (eq.n~\eqref{rho2}), starting from the initial conditions \eqref{in.cond.rho}, is bounded in any case, and vanishes asymptotically with $\sqrt{\beta(t)}$ (eq.n~\eqref{beta(t)}), in the presence of dissipation.

\begin{figure}[htbp]
\begin{center}
\includegraphics[width=3in]{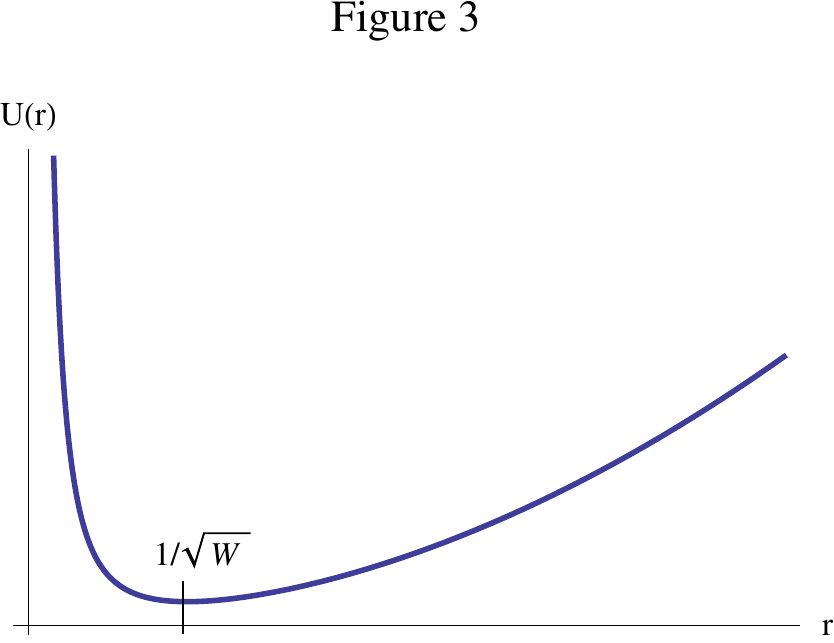}
\caption{\textbf{Instantaneous potential energy of the oscillator eq.n~\eqref{eq.r}}.}
\label{default}
\end{center}
\end{figure}

Let, in contrast, $q$ oscillate with diverging amplitudes, due to PR. Since $I\ge(q/\rho\sqrt{2})^2$ is an invariant, equation \eqref{I} shows that $\rho$ diverges at least as $|q|$, in all the successions $\{t^*_n\}$ such that $|q(t^*_n)|\rightarrow\infty$. Actually, around those points the repulsive force in eq.n~\eqref{eq.rho} is vanishing small, asymptotically, which makes eq.n~\eqref{eq.rho} coincide with eq.n~\eqref{ECl}. In conclusion: 

\begin{equation*}
\mathrm{lim}_{n\rightarrow\infty}|q(t^*_n)|=\infty\quad\Rightarrow\quad\mathrm{lim}_{n\rightarrow\infty}\rho(t^*_n)=\infty\:,
\end{equation*}
\\ 
which shows that PR transfers itself from $q$ to $\rho$ through the invariant $I>0$.

\section{Discusssion and Conclusions}
\label{concl}

Parametric resonance (PR) is a special case of wide applicatibility, whose main effect is the exponential divergence of a \emph{classical} oscillator's energy at long times, even in the presence of damping:

\begin{equation}
\label{Ecl2}
E_{cl}(t)\rightarrow E_{cl}(0)\mathrm{e}^{\Omega_{cl}t}\quad(\Omega_{cl}>0)\:.
\end{equation}
\\
The Kr\"{o}nig-Penney model (KPM) analyzed in the present work refers to a harmonic oscillator with piece-wise constant frequency $\widetilde{\omega}(t)$ and dissipation rate $\gamma(t)$, both time-changing in phase and periodically (Fig.1). The advantage of KPM is that the classical motion equation \eqref{ECl} can be solved exactly in terms of a transfer matrix $\mathcal{T}$, whose eigenvalues determine the PR bands, i.e. the intervals of values of $\omega_0\tau$ (the unperturbed frequency, times the period of the perturbation), in which PR prevails. As shown elsewhere \cite{Me}, the dissipationless model yields an infinite number of PR bands, extending themselves to arbitrary large values of $\omega_0\tau$. In Section \ref{CL}, this result is obtained as a special case, by showing that dissipation suppresses the high-frequency PR bands, leaving just a finite number, at most (Fig. 2).

The research of a motion invariant $I(p,\:q,\:\rho(t))$, different from the energy (eq.n~\eqref{I}), is an important issue, in the PR theory. Such invariant, however, requires an auxiliary function of time $\rho(t)$, satisfying an appropriate equation (eq.n~\eqref{eq.rho}). In Section \ref{q^-3}, the case $I=0$, is studied in detail, showing that $\rho(t)$ can be chosen in the asymptotic form:

\begin{equation}
\label{rho(A)}
\rho(t)=\sqrt{\beta(t)}\left[\frac{1}{\sqrt{\omega_0}}+\mathrm{o}(\delta(t),\:\xi(t))\right]\:,
\end{equation}
\\
where $\mathrm{o}(\delta(t),\:\xi(t))$ is first-order small in $\delta$ and $\xi$. This means that $\rho(t)$ is bounded anyway, and vanishes asymptotically in the presence of dissipation ($\beta(t)\rightarrow0$). If, in contrast, $I>0$ and $q(t)$ oscillates with diverging amplitudes (due to PR), then $\rho(t)$ oscillates with diverging amplitudes too, scaling with those of $|q(t)|$.

As it stands, the present work is just a pedagogical issue, solving a fairly complicated problem of classical mechanics, but without special new insights. However, in the author's intention the present work is also a preliminary approach to a more advanced question: the \emph{quantum} evolution of an oscillator with periodically time-varying frequency and damping. In particular, the different behaviors of the classical auxiliary function $\rho(t)$, exploited in Section~\ref{q^-3}, turn out to have a paramount importance in the quantum case \cite{DEFL}, as will be shown in a forthcoming paper.

\begin{appendices}
\numberwithin{equation}{section}
\section{Appendix A}

Since now on, let $\mathrm{o}^j$ indicate any quantity small to order $j$ in $\delta$ and/or $\xi_{0,1}$. For the quantites $A$ and $R$ (eq.ns~\eqref{A}, eq.n~\eqref{lambda2}), the lowest order approximation yields:

\begin{align}
A&=1+\frac{\delta^2}{8}+\frac{\Delta\xi^2}{2}+\mathrm{o}^3\nonumber\\
\nonumber\\
R&=\cos(2\langle\:\theta\:\rangle)-\left(\frac{\delta^2}{8}+\frac{\Delta\xi^2}{2}\right)\sin\left(\langle\:\theta\:\rangle+\frac{\Delta\theta}{2}\right)\sin\left(\langle\:\theta\:\rangle-\frac{\Delta\theta}{2}\right)+\mathrm{o}^3=\nonumber\\
\nonumber\\
&=\cos(2\langle\:\theta\:\rangle)+\left(\frac{\delta^2}{8}+\frac{\Delta\xi^2}{2}\right)\left[\sin^2(\Delta\theta/2)-\sin^2(\langle\:\theta\:\rangle)\right]+\mathrm{o}^3\:.\label{R}
\end{align}
\\
From eq.n~\eqref{R}, it follows immediately that $|\cos(2\langle\:\theta\:\rangle)|$ must be close to 1, in order that $R^2>1$, since the second term in the r.h.s. is of order $\mathrm{o}^2$. This preliminary condition for PR then reads (recall eq.n~\eqref{thetas}):

\begin{equation}
\label{<theta>}
2\langle\:\theta\:\rangle=\omega_0\tau\left[1-\frac{\delta}{4}\left(1-\frac{\Delta t}{\tau}\right) +\mathrm{o}^2\right]=n\pi+\phi\quad\Rightarrow\quad\Delta\theta=n\pi\frac{\Delta t}{\tau}+\mathrm{o}^1\:,
\end{equation}
\\
with $n=1,\:2,\:\cdots\:$ and $|\phi|<<1$, of the same order as $\delta$ and $\xi_{0,\:1}$. Equation \eqref{omegatau} follows from the first formula \eqref{<theta>}. Both expressions \eqref{<theta>}, inserted in eq.n~\eqref{R} yield:

\begin{equation}
\label{R2}
R=(-1)^n\left(1- \frac{\phi^2}{2}\right)+\left(\frac{\delta^2}{4}+\Delta\xi^2\right)\left[\sin^2\left(\frac{n\pi\Delta t}{2\tau}\right)-\sin^2\left(\frac{n\pi}{2}\right)\right]+\mathrm{o}^3\:.
\end{equation}
\\
On noticing that

\begin{align*}
&\sin^2\left(\frac{n\pi\Delta t}{2\tau}\right)-\sin^2\left(\frac{n\pi}{2}\right)=
\begin{cases}
\sin^2\left(\frac{n\pi\Delta t}{2\tau}\right)\:&\quad(n \text{ even})\\
\\
-\cos^2\left(\frac{n\pi\Delta t}{2\tau}\right)\:&\quad(n \text{ odd})
\end{cases}=\nonumber\\
\nonumber\\
&=\frac{(-1)^n-\cos^2(n\pi\Delta t/\tau)}{2}\:,
\end{align*}
\\
from eq.n~\eqref{R2} one gets:

\begin{align}
&R\pm\sqrt{R^2-1}=\nonumber\\
\nonumber\\
&=(-1)^n\pm\sqrt{\left(\frac{\delta^2}{4}+\Delta\xi^2\right)\frac{1-(-1)^n\cos^2(n\pi\Delta t/\tau)}{2}-\phi^2}\label{R3}\:,
\end{align}
\\
which shows that $\lambda_+$ (first eq.n~\eqref{lambda3}) can make PR prevail for $n$ even, and $\lambda_-$ does the same for $n$ odd. In both cases, the condition $|\lambda_\pm|>1$ (eq.n~\eqref{lambdaall2}) yields, to lowest order:

\begin{equation*}
\widetilde{\gamma}\tau=n\pi\left(\langle\:\xi\:\rangle+\frac{\Delta\xi\Delta t}{2\tau}\right)+\mathrm{o}^2<\ln\left[\left|R\right|+\sqrt{R^2-1}\right]\:,
\end{equation*}
\\ 
according to the second eq.n~\eqref{lambda3} and to eq.n~\eqref{omegatau}. With the aid of eq.n~\eqref{R3}, the preceding inequality finally leads to the bandwidth expression eq.ns~\eqref{Bandwidth}, with $\phi:=\omega_0\tau-(\omega_0\tau)_n$.

\end{appendices}

\end{document}